\documentclass{article}

\usepackage{arxiv}

\usepackage[utf8]{inputenc} 
\usepackage[T1]{fontenc}    
\usepackage{hyperref}       
\usepackage{url}            
\usepackage{booktabs}       
\usepackage{amsfonts}       
\usepackage{nicefrac}       
\usepackage{microtype}      
\usepackage{lipsum}
\usepackage{graphicx}
\usepackage{authblk}
\usepackage{gensymb}
\usepackage{float}
\graphicspath{ {./images/} }
\usepackage{mathtools}
\usepackage{amsmath,amssymb}
\usepackage{comment}
\usepackage{mathrsfs}
\usepackage{graphicx,nicefrac}
\usepackage{float}
\usepackage{amssymb}

\usepackage{comment}
\usepackage{scalerel}
\usepackage[overload]{empheq}
\usepackage{textgreek}
\DeclarePairedDelimiter\abs{\lvert}{\rvert}%
\DeclarePairedDelimiter\norm{\lVert}{\rVert}%

\providecommand{\va}{\mathbf{a}}

\providecommand{\vx}{\mathbf{x}}
\providecommand{\vy}{\mathbf{y}}
\providecommand{\vzz}{\mathbf{z}}
\providecommand{\vq}{\mathbf{q}}
\providecommand{\vk}{\mathbf{k}}
\providecommand{\vrr}{\mathbf{r}}

\providecommand{\vu}{\mathbf{u}}
\providecommand{\vg}{\mathbf{g}}
\providecommand{\vu}{\mathbf{u}}
\providecommand{\vk}{\mathbf{k}}
\providecommand{\vp}{\mathbf{p}}
\providecommand{\vq}{\mathbf{q}}

\providecommand{\mJ}{\mathbf{J}}

\providecommand{\mD}{\mathbf{D}}
\providecommand{\mF}{\mathbf{F}}

\providecommand{\mJ}{\mathbf{J}}
\providecommand{\mP}{\mathbf{P}}
\providecommand{\mO}{\mathbf{O}}
\providecommand{\mH}{\mathbf{H}}
\providecommand{\mQ}{\mathbf{Q}}

\providecommand{\R}{\ensuremath{\mathbb{R}}}
\providecommand{\C}{\ensuremath{\mathbb{C}}}

\providecommand{\B}{\ensuremath{\mathbb{B}}}

\usepackage{algorithm}
\usepackage{algpseudocode}
\usepackage{enumitem}

\usepackage{mathrsfs}
\usepackage{graphicx,nicefrac}

\usepackage[utf8]{inputenc}
\usepackage[english]{babel}

\usepackage[margin=1cm]{caption}
\usepackage{comment}
\usepackage{scalerel}
\usepackage[overload]{empheq}

\usepackage{float}

\usepackage{mathtools}

\DeclareMathOperator*{\argmax}{arg\,max}

\usepackage{authblk}
\title{Quantitative Jones matrix imaging using vectorial Fourier ptychography}

\author[1]{Xiang Dai}
\author[1]{Shiqi Xu}
\author[1]{Xi Yang}
\author[1]{Kevin C. Zhou}
\author[2]{Carolyn Glass}
\author[1]{Pvan Chandra Konda}
\author[1,*]{Roarke Horstmeyer}

\affil[1]{Department of Biomedical Engineering, Duke University, Durham, NC 27708}
\affil[2]{Department of Pathology, Duke University, Durham, NC 27708}

\affil[*]{Corresponding author: roarke.w.horstmeyer@duke.edu}

\begin{document}
\maketitle
\begin{abstract}
This paper presents a microscopic imaging technique that uses variable-angle illumination to recover the complex polarimetric properties of a specimen at high resolution and over a large field-of-view. The approach extends Fourier ptychography, which is a synthetic aperture-based imaging approach to improve resolution with phaseless measurements, to additionally account for the vectorial nature of light. After images are acquired using a standard microscope outfitted with an LED illumination array and two polarizers, our vectorial Fourier Ptychography (vFP) algorithm solves for the complex 2x2 Jones matrix of the anisotropic specimen of interest at each resolved spatial location. We introduce a new sequential Gauss-Newton-based solver that additionally jointly estimates and removes polarization-dependent imaging system aberrations. We demonstrate effective vFP performance by generating large-area (29 mm$^2$), high-resolution (1.24 $\mu$m full-pitch) reconstructions of sample absorption, phase, orientation, diattenuation, and retardance for a variety of calibration samples and biological specimens. 
\end{abstract}

\section*{Introduction}

\label{sec:intro}
Polarization imaging methods provide a useful means to access information about the molecular arrangement of anisotropic samples in a label-free manner \cite{chipman2018polarized}, with applications spanning pathology \cite{pathpol,jin2003imaging,manjunatha2015histopathological,he2015characterizing,zhang2016wide,jan2015polarization,badreddine2016real,he2021polarisation}, developmental biology \cite{biopol}, and mineralogy \cite{panwar2020review}. As such, polarization is an important and exciting intrinsic contrast mechanism in microscopy, especially in combination with phase contrast and fluorescence imaging approaches \cite{guo2020revealing}. However, polarimetric imaging over wide areas at high resolution remains an outstanding challenge. Current microscopes that image at micrometer-level detail can observe a field-of-view (FOV) that covers several square millimeters~\cite{Zheng:14}, which makes polarimetric imaging of large specimens difficult. To overcome this challenge, most existing approaches scan in a step-and-repeat fashion to create large-area image composites \cite{jan2017collagen}. As a fully quantitative polarimetric recording typically requires multiple snapshots per measurement, this scanning process can be tedious and time-consuming. What’s more, it is also challenging to obtain phase-sensitive polarimetric measurements (e.g., retardance) with relatively simple microscope hardware, to step-and-scan while maintaining accurate phase sensitivity, and to minimize the impact of polarization-dependent properties of the imaging system on the acquired measurements.

\begin{figure*}[t!]
\begin{center}
  \includegraphics[width=15cm]{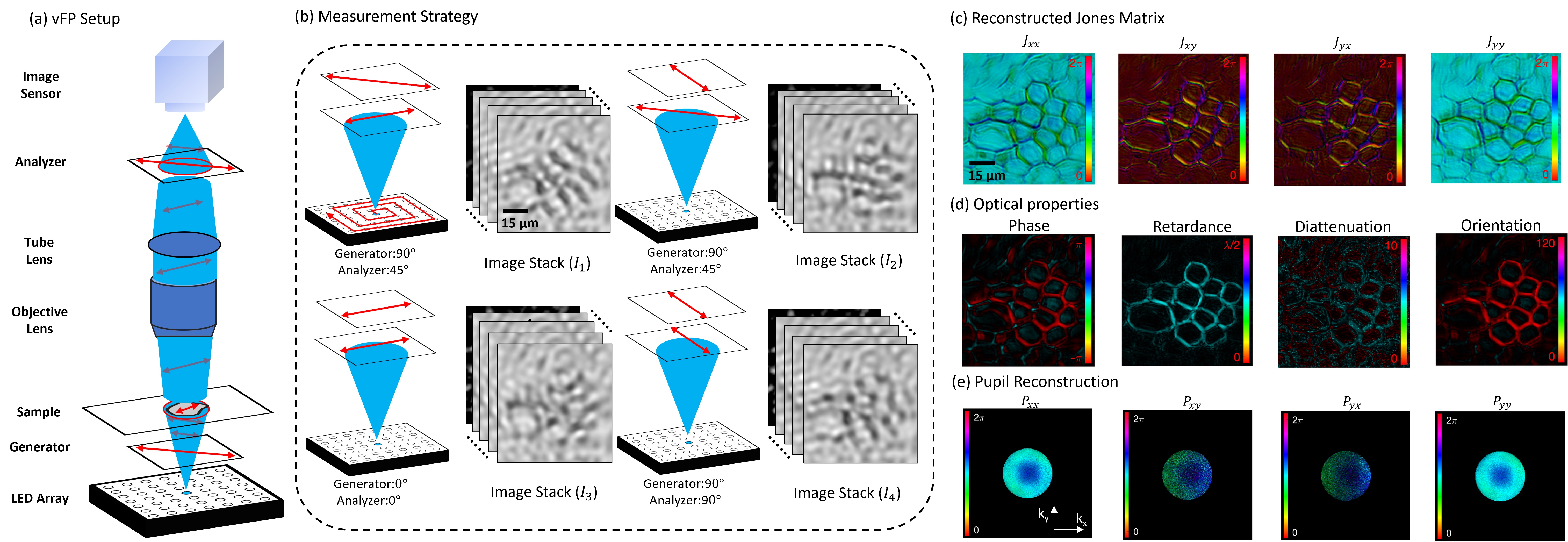} 
   \caption{Overview of vectorial Fourier ptychography (vFP). (a) vFP optical setup consists of a standard polarized light microscope modified with an LED array source to sequentially illuminate specimens with light polarized by generator. Light scattered in polarization-dependent manner passes through an analyzer before being recorded by the image sensor. (b) Images are acquired for each LED under four different generator-analyzer configurations. (c) 4-channel Jones Matrix vFP reconstruction of the cross-section of a broad bean root (color shows phase). (d) Various polarization-dependent sample properties extracted from the Jones matrix using eigenvector analysis. (e) vFP simultaneously measures a polarization-dependent pupil function as it removes associated per-channel aberrations. }
\label{teaser_fig}
\end{center}
\end{figure*}

In this work, we present a new approach for phase-sensitive polarimetric imaging that can address the above challenges and which produces high-resolution, wide area polarimetric measurements using relatively simple hardware. Our approach effectively extends an existing computational imaging method, termed Fourier ptychography~\cite{zheng2013wide}, to account for the vectorial nature of light during high resolution reconstruction. This extension, termed vectorial Fourier ptychography (vFP), first acquires a sequence of intensity images under variable-angle illumination provided by an LED array. It then utilizes a phase retrieval algorithm to recover complex polarimetric image reconstructions, summarized by a complex-valued Jones Matrix at each resolved pixel, from which one can directly measure quantities such as specimen retardance, orientation and diattenuation. The resolution of these reconstructions surpasses the standard resolution limit defined by the microscope objective lens: here, we use a 4X 0.1 NA objective lens and $f=100mm$ tube lens to provide 2X overall system magnification (29 $mm^2$ FOV), with an expected two-point resolution limit of 6.32$\mu$m. We reconstruct complex-valued images with a final two-point resolution of 1.24$\mu$m, typically associated with 10X-20X objective lens performance. Our computational strategy also jointly estimates and removes the imaging system’s polarization-dependent aberrations, which points the way towards a new class of simple and relatively inexpensive microscopes that can produce high-quality and quantitative complex polarimetric measurements over large areas without step-and-repeat scanning. 

In the remainder of this paper, we will first review existing approaches for polarimetric imaging, then present the forward and inverse mathematical models for vectorial Fourier ptychography, before experimentally demonstrating its effective quantitative performance. 

\section{Background}
There are three general categories of polarization-sensitive microscopic imaging. The first category captures qualitative polarimetric images, including the use of differential interference contrast (DIC), and placing a specimen between two polarizing films to visually assess variations of transmitted light~\cite{bass2009handbook}. To obtain quantitative polarimetric measurements, a second category of approach relies upon incoherent illumination to measure a specimen’s Muller matrix~\cite{pezzaniti1995mueller}, which describes the specimen’s optical response via a 4x4 real-valued matrix at each spatial location of interest. In this class of incoherent approaches, it is typical to describe light via a 4-entry Stokes vector, and a complete specimen reconstruction typically requires at least sixteen measurements per imaged pixel~\cite{goldstein2017polarized}. There have been a variety of optical geometries developed for full Mueller matrix-based polarization measurement ~\cite{hielscher1997diffuse,yao1999two,chipman2018polarized,goldstein2017polarized,he2021polarisation}. By adopting circularly polarized specimen illumination and applying certain specimen approximations, various setups can obtain fewer image measurements to obtain specimen properties like birefringence and orientation\cite{he2021polarisation}.

A third category of quantitative polarimetric microscopy relies upon coherent light and can be described using the Jones Matrix formalism \cite{jones1941new}. Within this category, the optical field is commonly modeled as two-dimensional complex vector, and associated optical transforms (including the interaction of light with thin specimens) as $2\times2$ complex matrix transforms. Typically, phase-sensitive measurements are needed to infer the $2\times2$ specimen Jones Matrix at each resolved location in the specimen. Prior work has considered this type of approach while detecting light with interferometry~\cite{wang2008jones,aknoun2018quantitative,jiao2020real,saba2021polarization,colomb2002polarization,kim2012polarization,yang2016single,ge2021single}, optical coherence tomography~\cite{de2017polarization,yao1999two}, and laser scanning confocal microscopy~\cite{massoumian2003quantitative}, for example. Unlike incoherent techniques, phase-sensitive methods enable direct measurement of specimen phase retardance for topological information, as well as specimen diattenuation, which is the complex differential transmittance within independent polarization channels.

To alleviate the complexity introduced by interferometric detection methods, a number of strategies have been recently developed that jointly utilize a post-processing algorithm to assist with recovering a specimen's complex polarization-dependent properties ~\cite{oldenbourg2013polarized,mehta2013polarized,shin2018reference,bai2020pathological,song2021large,song2020ptychography,yeh2021upti}. These computational imaging-based techniques offer a variety of unique benefits, such as improved imaging speed and reduced system complexity. In this work, we achieve complex-valued Jones matrix imaging at high resolution over a large field-of-view by using principles from Fourier ptychography (FP)~\cite{zheng2013wide}. Recent work has considered how FP can improve polarimetric imaging~\cite{song2021large}, but did not demonstrate the ability to recover the specimen's complex-valued Jones matrix, like many of the interferometric approaches above.  Complex Jones matrix reconstruction offers a complete picture of the polarimetric properties of thin specimens and has been demonstrated in scanning-based ptychography~\cite{ferrand2018quantitative,song2020ptychography}, which uses a completely different measurement arrangement but is mathematically similar to FP \cite{horstmeyer2014phase,konda2020fourier}. In this work, we build upon these prior demonstrations in ptychography to jointly reconstruct quantitative, complex-valued Jones matrix measurements at high-resolution over a large area, along with the additional benefit of computationally accounting for polarization-dependent imaging system aberrations.

\section{Methods}
\label{sec::method}

\subsection{vFP forward model}
\label{sec::forward model}

We begin by developing a mathematical expression for Fourier ptychographic imaging using Jones calculus, which will allow us to describe our data collection strategy using vectorial optical fields. A diagram of the vFP optical setup is in Fig.~\ref{teaser_fig}(a). Jones calculus represents an optical field as a $2\times1$ complex vector containing two orthogonal components, which we define here along the $x$ and $y$ axes. The light's polarization state is defined by the amplitude and phase of these two components at each spatial location of interest. General optical transformations (e.g., passage through an optical component) are summarized by $2\times2$ complex matrices. We note that the Jones calculus employed here does not directly consider depolarizing effects, or 3D effects, for example, which can easily be added in at the expense of a more complex model. 

Our setup consists of a programmable LED array as illumination source, a generator polarizer placed between the sample and the LED array, standard microscope optics, a CMOS camera, and an analyzer polarizer placed between the optics and camera. The $n$th tilted plane wave emitted by the $n$th LED within the array is described by $\text{exp}(i\vk_n \cdot \vrr)$, which is polarized by the generator to form a $2 \times 1$ vector field, $\mathbf{G_n^m }=\left[g_{x}^m\text{exp}(i\vk_n\cdot\vrr), g_{y}^m\text{exp}(i\vk_n\cdot\vrr)\right]^T$, which is obliquely-incident on the sample of interest. Here, $m\in\{0^{\circ},45^{\circ},...\}$ specifies a particular polarization state, and we use $\mathbf{g_m} = \left[g_{x}^m, g_{y}^m\right]$ as complex-valued scalars to define the appropriate generator-dependent weighting for each vector component. The polarized plane wave $\mathbf{G_n^m }$ then interacts with the thin specimen of interest, which we model as a product with a $2 \times 2$ complex matrix $\boldsymbol{\bar{O}}(\vrr)$. 

The resulting field then propagates to the imaging system's back focal plane, which we model (under certain approximations) with a $2 \times 2$ matrix $\mathbf{F}$ that contains the two dimensional Fourier transform operator $\mathcal{F}$ along its diagonal. Scalar FP models imaging system aberrations with a scalar pupil function $P(k)$~\cite{Ou:14}. In this work, we extend this model to also allow for polarization-dependent aberrations with a $2 \times 2$ complex pupil function matrix $\mathbf{P}(\mathbf{k})$, which accounts for both standard aberrations (e.g., defocus, astigmatism) and more complex birefringent effects, for example as introduced by plastic lenses. 

After passage through the pupil function (i.e., multiplication with $\mathbf{P}(\mathbf{k})$), the vectorized field then propagates to the image plane, which we model here with a $2 \times 2$ (inverse) Fourier transform matrix, although alternative transforms can be directly inserted. The vector field finally passes through a second ``analyzer" polarizer, $\mathbf{A_{l}}$, $(l\in\{0^{\circ},45^{\circ},\mathrm{right\ circular,}...\})$, to produce the  $n$th vector field $\mathbf{E_n^{(l,m)}}$ at the image plane,

\newcommand{\Omatrix}{
  \begin{bmatrix}
   &\mkern-18mu O_{xx}(\vk) &O_{xy}(\vk) \\
   &\mkern-18mu O_{yx}(\vk) &O_{yy}(\vk) 
   \end{bmatrix}}
\newcommand{\OmatrixSpace}{
  \begin{bmatrix}
   &\mkern-18mu \bar{O}_{xx}(\vrr) &\bar{O}_{xy}(\vrr) \\
   &\mkern-18mu \bar{O}_{yx}(\vrr) &\bar{O}_{yy}(\vrr) 
   \end{bmatrix}}
\newcommand{\Pmatrix}{
  \begin{bmatrix}
   &\mkern-18mu P_{xx}(\vk) &P_{xy}(\vk) \\
   &\mkern-18mu P_{yx}(\vk) &P_{yy}(\vk) 
   \end{bmatrix}}
\newcommand{\Fmatrix}{
  \begin{bmatrix}
   &\mkern-18mu \mathcal{F}^{-1}_{2D} & 0 \\
   &\mkern-18mu 0 & \mathcal{F}^{-1}_{2D} 
   \end{bmatrix}}

\begin{equation} 
    \mathbf{E_n^{(l,m)}}=\mathbf{A}_l\mathbf{F}^{-1} \mathbf{P}(\mathbf{k})\mathbf{F}\mathbf{\bar{O}}(\mathbf{r})\mathbf{G_n^m}
\end{equation}
where 
\begin{equation}
   \bar{\mO}(\vrr) = \OmatrixSpace, \mP(\vk) = \Pmatrix
\end{equation}
are the Jones matrices of the spectrum of the sample and pupil. We can take advantage of the Fourier shift theorem to remove the first Fourier transform matrix, and instead describe a shifted sample spectrum across the pupil plane, caused by each plane wave traveling at angle $\vk_n$, as $\mathbf{O}(\mathbf{k}-\vk_n)$, where $\mathbf{O}$ is the Fourier transform of $\mathbf{\bar{O}}$. This leads to the following compact form of the vFP forward model for each detected intensity image $i$ under illumination from the $n$th LED angle with generator and analyzer configurations $(l,m$): 

\begin{equation}
    i_{n}^{l,m}\left(\mathbf{O},\mathbf{P}\right)=\abs{{\va_l}^T\mF^{-1}\mP(\vk)\mO(\vk-\vk_n)\vg_m}^2,
\label{eq:forward_general}
\end{equation}
where $\va_l$ and $\vg_m$ is the Jones vector of the $l^{th}$ analyzer and $g^{th}$ generator, respectively. In our first demonstration we use linear polarizers for both generator and analyzer, as will be discussed in the next subsection, we have $\va_l=\left[\cos{\alpha_l},\sin{\alpha_l}\right]^T$, and $\vg_m=\left[\cos{\gamma_m},\sin{\gamma_m}\right]^T$, where $\alpha_l$ and $\gamma_m$ are the angles of the linear polarizers of the $l^{th}$ analyzer and $g^{th}$ generator, respectively. The goal of vFP is to jointly recover the complex $2 \times 2$ matrices describing the specimen and pupil function, $\boldsymbol{O}$ and $\mathbf{P}$, from an acquired image dataset $i_n^{l,m}$ for multiple illumination angles $n=1$ to $N$ and multiple configurations of the generator $l$ and the analyzer $m$. There are various possible selections for these measurement variables, and we next describe one effective measurement strategy that yields a well-conditioned image dataset for inverse problem estimation.

\subsection{Measurement strategy}
\label{sec::measStrat}
Obtaining four unique measurements per illumination angle is one effective strategy to solve for the four unknowns summarized by the sample matrix $\mO(\vk)$. There are, of course, many possible configurations of the generator and analyzer to utilize for such measurements\cite{ferrand2018quantitative}. Specific selections significantly reduce the complexity of Eq.~ (\ref{eq:forward_general}). For example, selecting the generator($G$) and the analyzer($A$) to produce and admit zero-degree linearly polarized light will result in $\vg_m=[1,0]^T$ and $\va_l=[1,0]^T$. The resulting image intensity will be described by $i_{n}^{l=0\degree,m=0\degree}=\abs{\mathcal{F}^{-1}[P_{xx}O_{xx}(\vk-\vk_n)+P_{xy}O_{yx}(\vk-\vk_n)]}^2$ for plane wave illumination at an angle $\vk_n$. Setting the generator and the analyzer at 90 degrees will likewise yield additional complementary information about $\mathbf{O}$ and $\mathbf{P}$. The above two configurations provide the helpful benefit of high optical transmission for most specimens of interest, which facilitates the use of lower exposure times and reduced noise. To balance the experimental benefits of high optical transmission with the computational benefits of measurement redundancy, we next set the generator ($G$) at $45\degree$ for our remaining two measurements. Our four polarization configurations were thus, $(l,m)\in\{G\mkern-6mu:0\degree, A\mkern-6mu:0\degree; G\mkern-6mu:90\degree,A\mkern-6mu:90\degree; G\mkern-6mu:45\degree,A\mkern-6mu:0\degree; G\mkern-6mu:45\degree,A\mkern-6mu:90\degree\}$, as shown in Fig.~\ref{teaser_fig}. We note that alternative generator/analyzer configurations are certainly possible and compatible with our forward and inverse solver, and will be the subject of future study.

\subsection{Inverse problem}
Using the measurement strategy proposed in Section \ref{sec::method}\ref{sec::measStrat}, the four amplitude measurements for the $n^{th}$ LED are 
\begin{subequations}
\label{eq:forward_eq}
{\footnotesize
    \begin{empheq}[left={\empheqlbrace\,}]{align}
     i_{n}^{0\degree,0\degree}(\vrr)  = \abs{\mathcal{F}^{-1}&\{P_{xx}(\vk)\cdot O_{xx}(\vk-\vk_n)+P_{xy}(\vk)\cdot O_{yx}(\vk-\vk_n)\}}^2\\
     i_{n}^{90\degree,90\degree}(\vrr)  = \abs{\mathcal{F}^{-1}&\{P_{yx}(\vk)\cdot O_{xy}(\vk-\vk_n)+P_{yy}(\vk)\cdot O_{yy}(\vk-\vk_n)\}}^2\\
     i_{n}^{45\degree,90\degree}(\vrr)  = \abs{\mathcal{F}^{-1}&\{P_{xx}(\vk)\cdot O_{xx}(\vk-\vk_n)+P_{xx}(\vk)\cdot O_{xy}(\vk-\vk_n)\nonumber\\ 
     +&P_{xy}(\vk)\cdot O_{yx}(\vk-\vk_n)+P_{xy}(\vk)\cdot O_{yy}(\vk-\vk_n)\}}^2\\
     i_{n}^{45\degree,90\degree}(\vrr)  = \abs{\mathcal{F}^{-1}&\{P_{yx}(\vk)\cdot O_{xx}(\vk-\vk_n)+P_{yx}(\vk)\cdot O_{xy}(\vk-\vk_n)\nonumber\\ 
     +&P_{yy}(\vk)\cdot O_{yx}(\vk-\vk_n)+P_{yy}(\vk)\cdot O_{yy}(\vk-\vk_n)\}}^2.
    \end{empheq}}
\end{subequations}
We have developed an iterative algorithm to estimate the complex-valued object and pupil matrices (extending scalar phase recovery~\cite{tian2014multiplexed,maiden2017further} to the vector case) by minimizing the Euclidean distance between predicted and measured amplitude,
\begin{equation}
\label{eq:datafit}
    \mathcal{D}(\mO,\mP)=\frac{1}{2 N}\sum_n^N\sum_{(l,m)}^{}\norm{\sqrt{i_{n}^{l,m}\left(\mathbf{O},\mathbf{P}\right)}-\sqrt{\hat{i}(\vrr)_{n}^{l,m}}}^2_2.
\end{equation}
Here, $N$ is the total number of illumination angles, while ${i}_n^{l,m}\left(\mathbf{O},\mathbf{P}\right)$ and $\hat{i}_n^{l,m}$ are the vFP forward model prediction (eq.\ref{eq:forward_eq}) and the actual experimental image measurements for the $n$th LED illumination using the $(l,m)$th generator-analyzer configuration. To minimize the loss function $\mathcal{D}(\mO,\mP)$, we use a Gauss-Newton-based sequential solver, which alternatively updates the object and pupil matrix estimations based upon the 4 polarization measurements from each LED. Our method is inspired by prior ptychographic phase retrieval solvers like the sequential Gauss-Newton(GN) method~\cite{tian2014multiplexed} and regularized ptychographic iterative engine (rPIE)~\cite{maiden2017further}, which use CR-calculus~\cite{kreutz2009complex} to estimate the gradient of the loss with respect to both sample and pupil. In our approach, we stack 4 image measurements into a per-LED measurement vector and also compute a complex-valued loss function gradient. Unlike prior work, minimizing the loss function with a GN method requires inversion of a non-diagonal system matrix due mixing of polarization states. This adds several complexities as compared to matching computations with scalar fields, which can effectively use a direct division for inversion~\cite{tian2014multiplexed,maiden2017further}. We refer the interested reader to a complete presentation and derivation of our algorithm in the Supplementary Material.

\begin{figure}[t!]
\begin{center}
    \includegraphics[width=8.5cm]{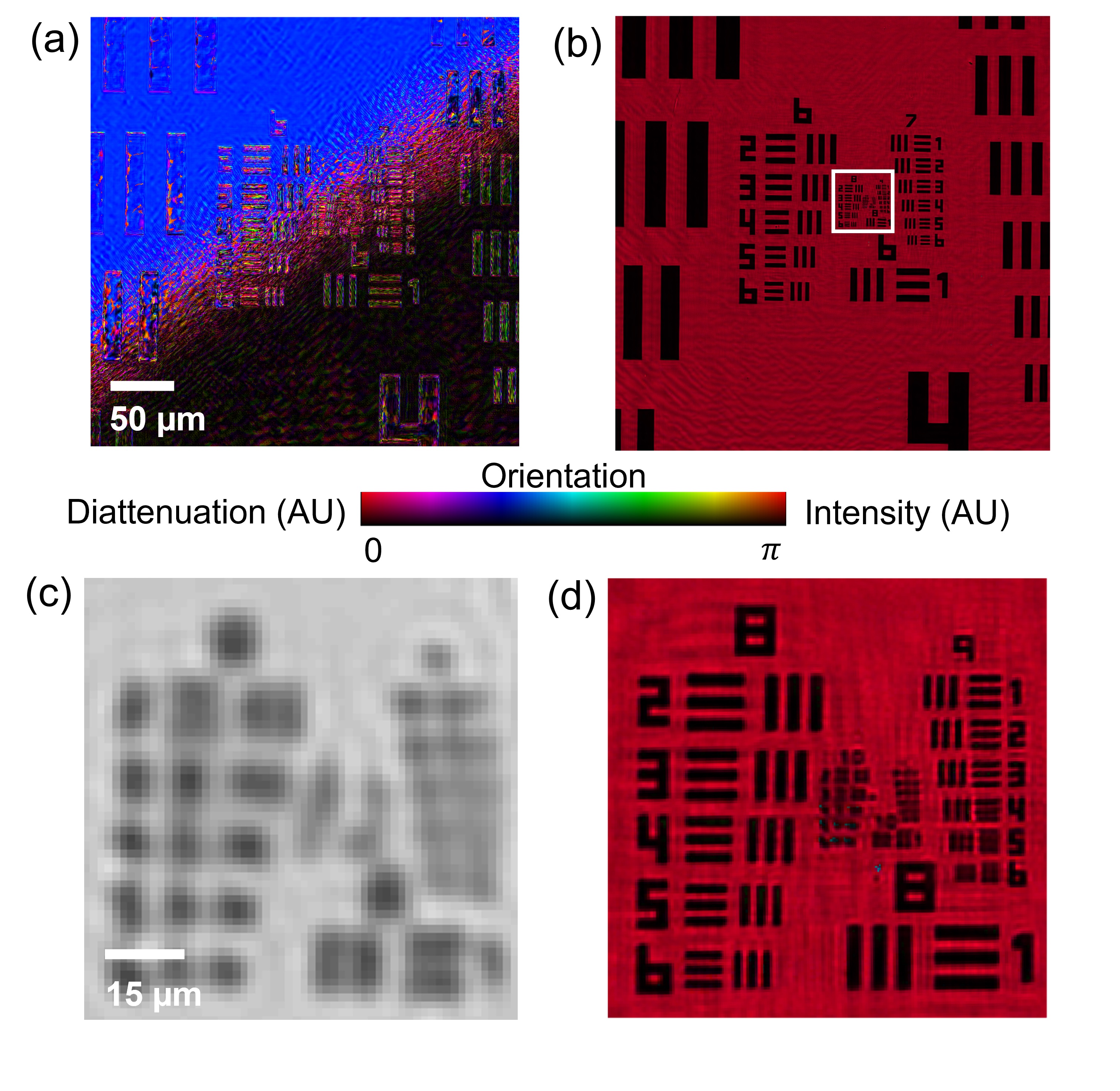}
    \caption{Verification of vFP diattenuation estimation accuracy, as well as resolution gain, using a resolution target and linear polarizer. (a) Diattenuation (brightness) and orientation (hue) of vFP reconstruction with USAF target half covered (in the upper left) by a linear polarizer. (b) Incoherent sum of squared channel amplitudes (brightness) and orientation (hue) of USAF target fully covered by the linear polarizer, rotated by 48 degrees as compared to placement in (a). (c) Zoom-in of the raw image captured under illumination from center LED. (d) Zoom-in of the vFP reconstruction in (b) verifies resolution gain.}
    \label{Fig.2}
\end{center}
\end{figure}

Once we reconstruct the $2\times2$ Jones matrix for the specimen and pupil, we use a simple eigendecomposition method~\cite{chipman2018polarized} to derive the homogeneous anisotropic properties of the specimen, including retardance, diattenuation and orientation. Specifically, for the 2x2 Jones matrix $\bar{\mO}(\vrr)$ per reconstructed pixel, we first compute eigenvalues $\xi_q$, $\xi_r$ and eigenvectors $\mathbf{E}_q$, $\mathbf{E}_r$. Under the assumption of homogeneous anisotropic specimen, the calculated eigenvectors indicate two perpendicular specimen axes with associated eigenvalues that can be used to compute a per-pixel retardance,
\begin{equation}
    B = \frac{\lambda}{2\pi}\abs{\angle\xi_q-\angle\xi_r},
\end{equation}
where $\angle$ denotes complex angle. Similarly, the two eigenvalues may be used to compute an amplitude-dependent diattenuation measurement:
\begin{equation}
    D=\Bigg\lvert\frac{\abs{\xi_q}^2-\abs{\xi_r}^2}{\abs{\xi_q}^2+\abs{\xi_r}^2}\Bigg\rvert.
\end{equation}
In this work, we adopt a definition for orientation as the angle between the reference axis given by the linear polarizer within the setup and the slow axis within the specimen plane, where the slow axis is defined by the direction of the eigenvector whose corresponding eigenvalue phase is the larger of $\xi_q$ and $\xi_r$. Mathematically, orientation to the slow axis $\omega$ can be computed as the angle of the eigenvector,

\begin{equation}
    \omega = \tan^{-1}(\nicefrac{{E}_{i,x}}{{E}_{i,y}}), \quad i= \argmax_{i\in\{q,r\}}\angle\xi_i,
\end{equation}
where $E_{\cdot,x}$ and $E_{\cdot,y}$ are the real $x$ and $y$ components of the eigenvector, respectively.

\begin{figure*}[t!]
\begin{center}
    \includegraphics[width=15cm]{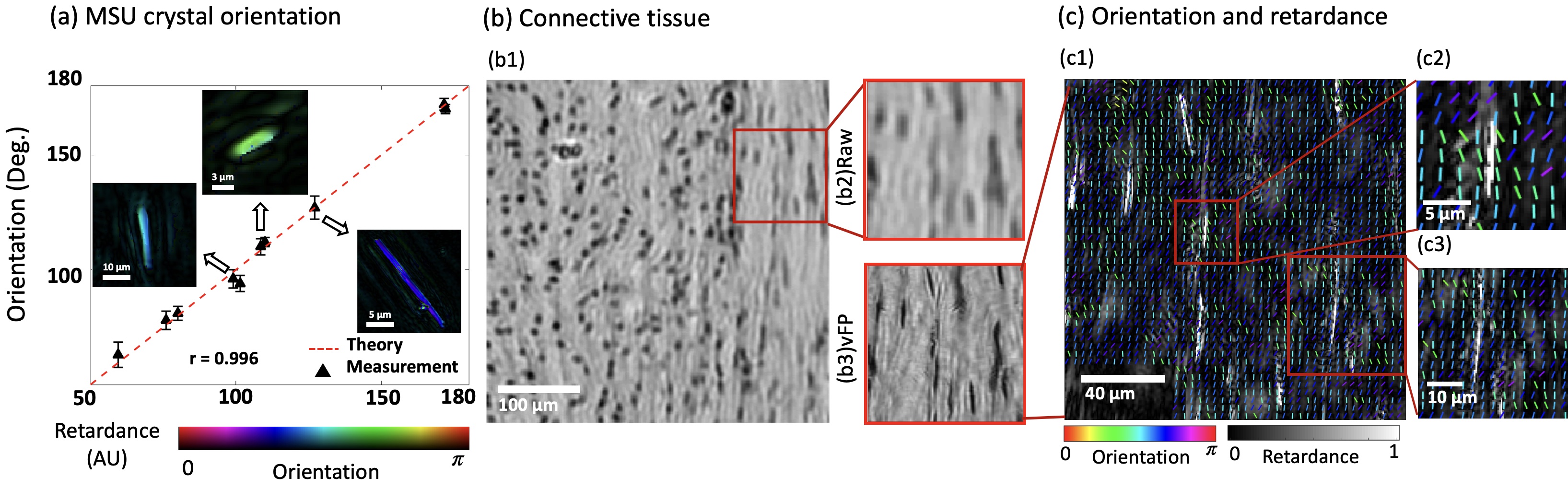} 
    \caption{Orientation measurement validation. (a) vFP orientation measurement of monosodium urate (MSU) crystals($y$-axis) exhibit expected linear relationship with crystal rotation angle ($x$-axis). Each point reports average and standard deviation (error bar) of per-crystal vFP orientation measurement averaged over pixels within each crystal (examples in 3 insets, color is orientation and intensity is retardance). (b) Raw image and vFP reconstruction of human connective tissue specimen. (b1) Example raw image and (b2) zoom-in. (b3) vFP reconstruction shown as incoherent summation of matrix elements. (c) vFP retardance and orientation of the same area as in (b2-3), where (c1) shows retardance in grayscale intensity and orientation false-colored and (c2, c3) highlight how specimen orientation follows fiber growth direction and exhibits spatial correlations with retardance.}
    \label{Fig.3}
\end{center}
\end{figure*}

\section{Results}
We verified our vFP model and inverse solver in a series of experiments using a standard microscope outfitted with an LED array. The vFP setup uses an objective lens (4$\times$, $NA_o$=0.1) and digital camera containing 2.4 $\mu m$ pixels (Basler acA5472-17um, 5496 × 3672 pixel count). The LED array contains 15 × 15 surface-mounted elements (center wavelengths: 632 nm, 525 nm and 470 nm per LED color with 4 mm LED pitch). We positioned the LED array 68 mm beneath the sample to create a maximum illumination NA of $NA_i=$ 0.41. This leads to an effective synthetic NA for our Fourier ptychographic reconstructions of $NA=NA_i+NA_o=0.51$, or a 5x improvement in spatial resolution in each lateral dimension. Linear wire grid polarizers (polarization wavelength range between 420 - 700 nm) were inserted directly before the LED array (generator) and before the camera (analyzer). Both linear polarizers were installed in rotation mounts with resonant piezoelectric motors (bidirectional accuracy, 0.002$^{\circ}$).

\subsection{Validation of vFP resolution and diattenuation}
To first validate that the polarization-dependent metrics of vFP exhibit the expected resolution enhancement commonly observed with scalar FP, we imaged a sample consisting of a plastic linear polarizer (Edmund Optics) placed above a high-resolution absorptive USAF target (Ready Optics). The linear polarizer provides a simple means to create an expected constant orientation and a uniform diattenuation across the regions it covers, while the USAF target provides a standard resolution measure.

The expected resolution gain of a vFP reconstruction follows the same principles as with scalar FP~\cite{Ou:15}, wherein the smallest resolvable full-pitch resolution (i.e., distance between the center of two similar bars) can be approximated by, $\frac{\lambda}{NA_{i}+NA_{o}}$. Using red LED illumination (center wavelength 632 nm) in this experiment, we accordingly expect a 1.24 $\mu$m full-pitch resolution cutoff after executing vFP, which corresponds to a cutoff slightly above USAF target Group 9 Element 5. The vFP orientation reconstruction achieves this resolution as shown in Fig. \ref{Fig.2}(d), which provides an appreciable gain over the raw captured image intensity in (c).

This linear polarizer experiment additionally provides a means to assess quantitative diattenuation and orientation measurement accuracy. As only the upper-left of the USAF target is covered by the polarizer in Fig.~\ref{Fig.2}(a), we expect a diattenuation value closer to 1 within this region, and close to 0 in the lower right, which is uncovered. The mean diattenuation and variance, averaged across a 500$^2$ pixel area within these two respective regions, are $(\bar{D}=0.9822, \sigma^2_D=0.001)$ and $(\bar{D}=0.0693, \sigma^2_D=0.011)$, respectively. We hypothesize that this experimentally measured diattenuation of less than 1 is due to the use of a plastic linear polarizer, whose transmission at 632 nm is close to 45\%, rather than 50\%. In addition, the transition boundary between the image area covered and uncovered by linear polarizer exhibits inaccurate values due to angle-dependent shadowing effects introduced by its finite thickness (i.e., the polarizer does not obey the required thin specimen approximation, leading to such artifacts). The diattenuation estimates at locations of absorptive USAF bars may also be inaccurate, as the bars are optically opaque. 

 \begin{figure}[t!]
\begin{center}
    \includegraphics[width=8.5cm]{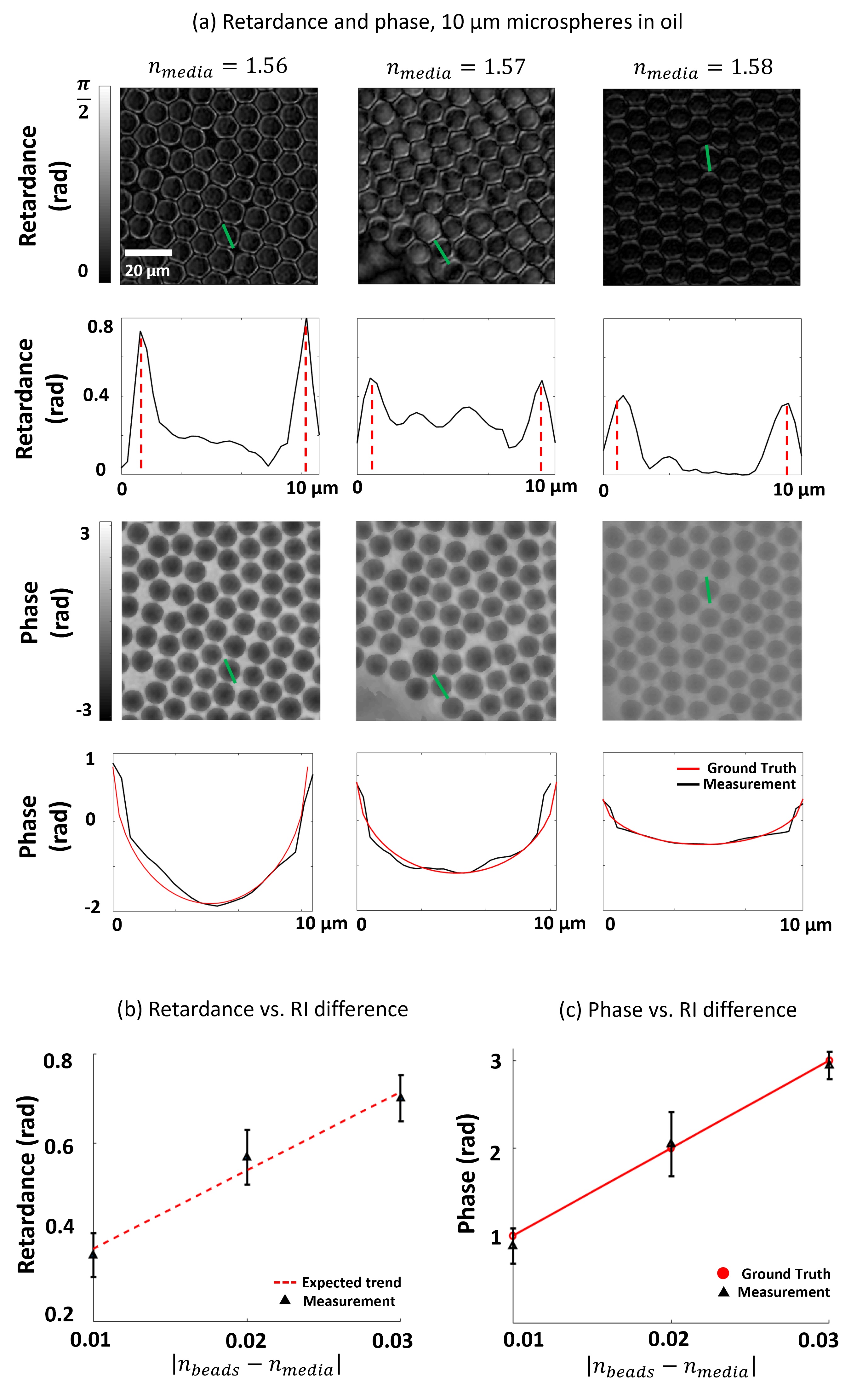}
    \caption{vFP retardance and phase validation experiments. (a) Retardance (top two rows) and phase (bottom two rows) reconstructions of 10-\textmu m monodisperse polystyrene microspheres immersed in oils with refractive indices (RIs) of 1.56, 1.57 and 1.58. Profiles through select beads show RI-dependent retardance shift at bead-oil boundary measured with red-dotted line, while phase profiles closely match ground truth shift. (b) Reconstructed maximum retardance shift (red dashed line in (a), average of 20 beads per point) follows expected linear trend with RI difference between microspheres and oil. (c) Reconstructed max. phase shift also follows ground truth (average of 20 beads per point).}
    \label{Fig.4}
\end{center}
\end{figure}

 \subsection{Quantitative orientation measurement}
 
 Orientation is a useful polarimetric quantity used within material and biological specimen analysis. To further investigate the quantitative accuracy of vFP for specimen orientation measurement, we next imaged a collection of randomly-oriented monosodium urate (MSU) crystals, which are are rod-like microstructures whose optical orientation properties are expected to exhibit a linear relationship with crystal rotation angle within the x-y plane. A summary of this experiment's results is presented in Fig. \ref{Fig.3}(a). Images of 3 example reconstructed MSU crystals are included as insets (color denotes reconstructed orientation and intensity denotes reconstructed retardance). Plotting the vFP-measured per-crystal orientation versus the expected orientation, as measured from crystal rotation angle within the x-y plane, reveals a clear linear relationship. Measured orientation points were averaged over all pixels within each crystal, where the error bar shows the resulting standard deviation. Ground-truth orientations were computed by fitting a line to each MSU crystal captured within a single FOV. 

In a final experiment to assess vFP's ability to measure orientation, we imaged a thin section of human connective tissue containing both muscle and collagen (Fig. \ref{Fig.3}(b-c)). In this type of tissue specimen, it is common to observe polarization-dependent effects due in part to the response of collagen fiber\cite{tuchin2016polarized,yang2021instant} To improve visualization, we overlay orientation as an undirected vector field on top of the retardance shown in grayscale in Fig. \ref{Fig.3}(c) to reveal a spatial correlation between retardance and orientation. Reconstructed orientation vector alignment is consistent with fiber growth direction within the connective tissue. We note the FOV in (c1) matches that of the example raw image is shown in (b2) and the zoom of the vFP reconstruction in (b3).

\begin{figure*}[ht!]
\centering\includegraphics[width=15cm]{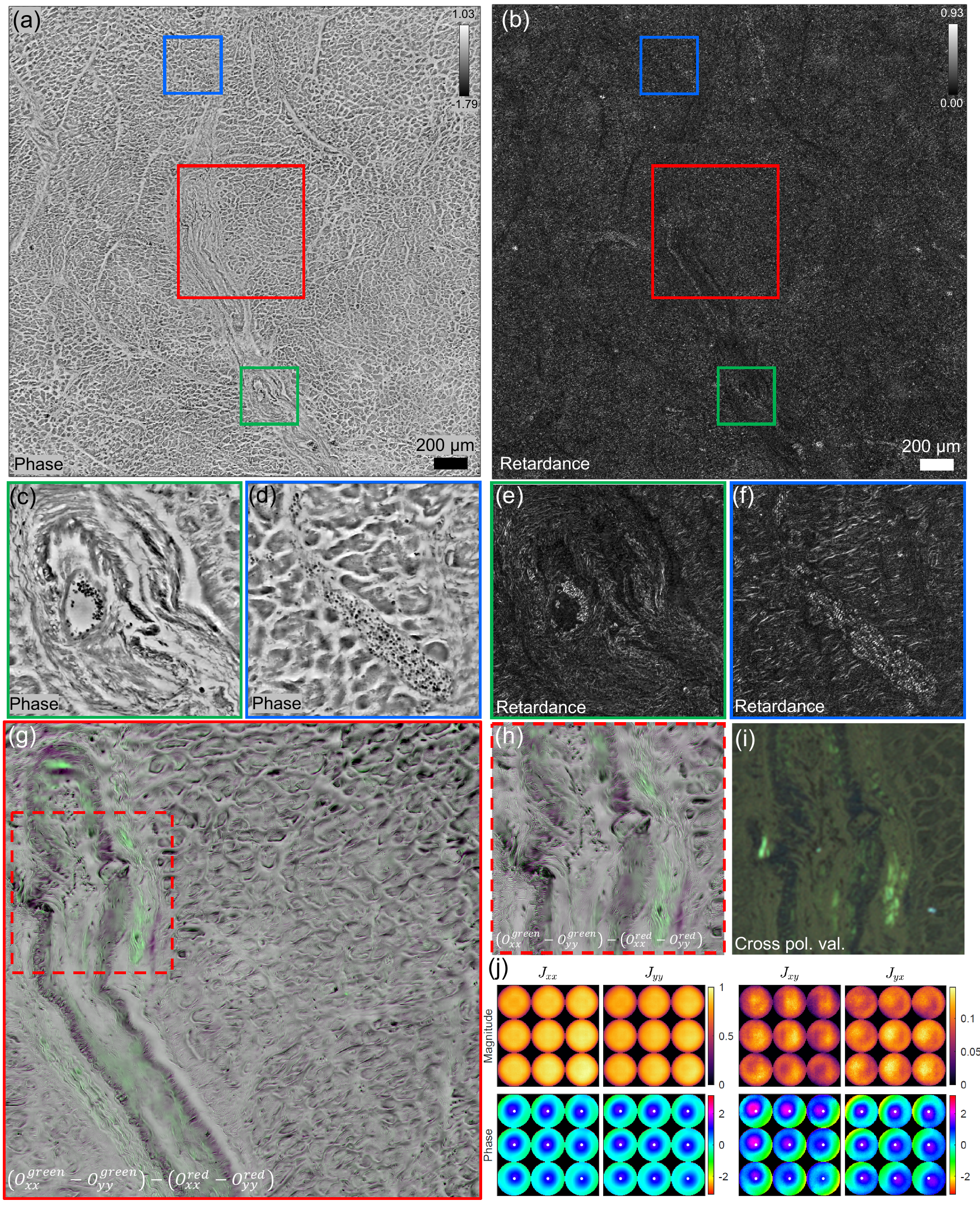}
    \caption{Large-FOV vFP reconstruction of thin cardiac tissue section stained with Congo red dye. (a) Phase reconstruction. (b) Retardance reconstruction. (c,d) Zoom-ins of (a). (e,f) Zoom-ins of (b). (g) Difference of the difference between the real $xx$ and $yy$ components of the Jones matrix reconstruction between red and green channels, corresponding to the red boxes in (a) and (b). (h) Zoom-in of (g). (i) Validation via bright-field microscope imaging with crossed polarizers. (j) Spatially-varying pupils, plotted as the magnitude and phase of the Jones matrix components. White dots denote the center of the pupil to improve visualization of shifts.
    }
    \label{Fig.5}
\end{figure*}
\subsection{Retardance and phase validation}

To validate vFP's ability to measure specimen retardance, we followed the findings of a recent study \cite{guo2020revealing} that demonstrated a linear relationship between retardance and refractive index (RI) difference at the boundary of microspheres and their embedding medium. Though microspheres do not intrinsically exhibit birefringence, a sharp RI transition at their boundary will lead to a measurable retardance shift that is linearly proportional to the RI difference at the transition boundary~\cite{oldenbourg1991analysis}. To explore this connection, we prepared 3 specimens of oil-immersed polystyrene microspheres with oils of 3 different RI to generate 3 unique RI transition differences. Images of vFP-reconstructed retardance are in Fig. \ref{Fig.4}(a) top, from which we observe a clear peak of retardance that consistently exists at the bead profile edge (i.e., forms a ring around each bead). Profiles of randomly selected beads below confirm such peak formation, from which we measure the maximum per-bead retardance shift (red dashed line). Plotting the average maximum retardance shift (20 beads per point) as a function of RI difference in Fig. \ref{Fig.4}(b) confirms the previously observed linear trend. Through this experimental analysis, we were additionally able to confirm the quantitative accuracy of vFP reconstructed phase values, as shown at the bottom of Fig.\ref{Fig.4}(a) and summarized in plot of average maximum phase shift versus RI difference in Fig. \ref{Fig.4}(c).

\subsection{Large-area amyloid plaque detection}
We next applied vFP to reconstruct the vectorial response of a specimen that is commonly viewed using polarization optics within the pathology lab: thin sections of fixed cardiac tissue stained with Congo red dye. Congo Red is known to stain specimen areas that contain amyloid fibrils in a manner that highlights birefringence as anomalous colors (typically, as an “apple green” shading) when viewed between two crossed polarizers using white light illumination in a transmission geometry \cite{howie2008physical}. Visual identification of such apple-green areas within the interstium and vessel walls of endomyocardial tissue is often an important diagnostic indicator of the presence of cardiac amyloidosis. In Fig.~\ref{Fig.5}(a,b), we present a large-FOV vFP reconstruction (8.27 $mm^2$ shown) of the phase and retardance of Congo red-stained cardiac tissue, wherein the arterial wall is clearly visible. Interestingly, in Fig. \ref{Fig.5}(f), the red blood cells (small, dark circular features) located within the vessels each exhibit a higher retardance along their border, which is consistent with our observations with polystyrene beads (Fig. \ref{Fig.4}). Examining the real component of the specimen Jones matrix entries $\bar{O}_{xx}$ and $\bar{O}_{yy}$ within this area shows minimal visual disparity under both green and red illumination wavelengths. However, by computing the difference ($\bar{O}_{xx} - \bar{O}_{yy}$) for each color channel, and subtracting the red channel's difference from the green, it is possible to produce a high-contrast map of Congo-red stained areas, as presented as a green overlay in Fig.~\ref{Fig.5}(f). Comparing the resulting channel difference map to what is typically observed with polarization optics in a standard microscope (white-light illumination, NA=0.25) demonstrates a close qualitative mapping. We hypothesize that vFP can thus provide a useful tool for examining the polarimetric response of such Congo-red-stained cardiac tissue over large areas at high resolution. 

Reconstructing a large FOV with vFP additionally allows us to examine how the recovered Jones matrix of the imaging system pupil function differs between on-axis and off-axis sample positions (Fig. \ref{Fig.5}(j)). Here, we divided the reconstruction FOV into $3\times3$ regions and averaged all pupils within each region. The Jones matrix pupil phase patterns experience a shift from the center for off-axis locations \cite{mcguire1994polarization}. Interestingly, the $J_{xy}$ and $J_{yx}$ pupil amplitudes are spatially varying, unlike $J_{xx}$ and $J_{yy}$. Higher-NA objectives or cheaper plastic lenses may exhibit stronger polarization effects, which will be the subject of future investigation.

\section{Discussion and Conclusion}

This paper proposes a new computational microscopy method, termed vectorial Fourier ptychography (vFP), to measure the complex amplitude and phase transmission of a thin sample along with its polarization properties contained within its complex Jones matrix. As a synthetic aperture imaging technique, vFP provides a means to obtain high-resolution Jones matrix measurements over a wide FOV without any complex moving parts. Through a simple modification of a conventional polarized light microscope (i.e., replacing the source with an LED array) and a new computational reconstruction algorithm, we demonstrated that our method can jointly recover a vectorial image of complex-valued 2x2 Jones matrices in a variety of calibrated phantoms and biological samples, along with estimation of the 2x2 Jones matrix summarizing the imaging system's vectorial response (i.e., its polarization-dependent aberrations). Our method thus exhibits capabilities that neither conventional FP nor polarized light microscopy can achieve alone.

As this initial demonstration was a proof of concept, there are several avenues for future direction. To improve acquisition speed, vFP can certainly borrow LED multiplexing \cite{tian2014multiplexed}, multispectral, \cite{loetgering2021tailoring} or multi-aperture \cite{konda2021multi} strategies previously demonstrated in conventional FP. Furthermore, the approach could also utilize a polarization-sensitive digital image sensor \cite{bai2020pathological,yeh2021upti}, which could reduce the amount of required captured data or otherwise improve vFP reconstruction fidelity. In addition, a screen display-based illumination mechanism could be adopted \cite{lee2021smartphone} to miniaturize the imaging system. While the present work was restricted to thin samples, additional theoretical development of the vFP framework may allow for 3D polarization microscopy \cite{saba2021polarization,yeh2021upti}in the future. Moreover, additional development may also account for the partial coherence of the LED illumination employed by vFP within the Stokes formalism \cite{yeh2021upti}. Finally, since vFP can jointly estimate and account for the polarization properties of the imaging system, our method could be applied to low-cost microscopes employing plastic lenses, which are known to exhibit birefringence effects under strain. Given the experimental simplicity of our method and the broad range of anisotropic biological samples, vFP could be widely adopted as a significant extension of conventional FP and polarized light microscopy.

\section{Funding}
This work was supported in part by a Duke-Coulter Translational Partnership Grant and a 3M Non-tenured Faculty Award.

\section{Acknowledgments}
We would like to thank Dr. Shalin Mehta, Dr. Ian Sigal, Dr. Bin Yang, Dr. Li-Hao Yeh, and Dr. Lars Loetgering for helpful guidance and feedback.

\section{Disclosures}
RH: Ramona Optics Inc. (F,I,P,S)

\bibliographystyle{unsrt}  
\bibliography{references}

\appendix
\section{A Sequential Gauss-Newton-based vectorial Fourier ptychography solver}
Using the configuration discussed in the \textit{Measurement strategy} section, the general vectorial Foruier ptychography(vFP) forward model can be simplified as 
\begin{subequations}
    \begin{empheq}[left={\empheqlbrace\,}]{align}
     {i_{1i}(\vrr)}  &= \abs{z_{1i}(\vrr)}^2\\
     {i_{2i}(\vrr)}  &= \abs{z_{2i}(\vrr)}^2\\
     {i_{3i}(\vrr)}  &= \abs{z_{3i}(\vrr)}^2\\
     {i_{4i}(\vrr)}  &= \abs{z_{4i}(\vrr)}^2,
    \end{empheq}
\end{subequations}
where
\begin{subequations}
    \begin{empheq}[left={\empheqlbrace\,}]{align}
     z_{1i}(\vrr)  = \mathcal{F}^{-1}\{P_{xx}(\vu)&\cdot O_{xx}(\vu-\vu_i)+P_{xy}(\vu)\cdot O_{yx}(\vu-\vu_i)\}\\
     z_{2i}(\vrr)  = \mathcal{F}^{-1}\{P_{yx}(\vu)&\cdot O_{xy}(\vu-\vu_i)+P_{yy}(\vu)\cdot O_{yy}(\vu-\vu_i)\}\\
     z_{3i}(\vrr)  = \mathcal{F}^{-1}\{P_{xx}(\vu)&\cdot O_{xx}(\vu-\vu_i)+P_{xx}(\vu)\cdot O_{xy}(\vu-\vu_i)\nonumber\\ 
     +&P_{xy}(\vu)\cdot O_{yx}(\vu-\vu_i)+P_{xy}(\vu)\cdot O_{yy}(\vu-\vu_i)\}\\
     z_{4i}(\vrr)  = \mathcal{F}^{-1}\{P_{yx}(\vu)&\cdot O_{xx}(\vu-\vu_i)+P_{yx}(\vu)\cdot O_{xy}(\vu-\vu_i)\nonumber\\ 
     +&P_{yy}(\vu)\cdot O_{yx}(\vu-\vu_i)+P_{yy}(\vu)\cdot O_{yy}(\vu-\vu_i)\}.
    \end{empheq}
\end{subequations}
Assuming everything is Nyquist rate sampled and stored as discrete signal, using vectorization notation, we can write the object as $\vx=[\vx_1;\vx_2;\vx_3;\vx_4]; \vx\in\C^{4m^2}$, where $\vx_1,\vx_2,\vx_3,\vx_4 \in\C^{m^2}$ are vectorization notations for  $O_{xy}(\vu),O_{xy}(\vu),O_{xy}(\vu),O_{xy}(\vu)$. $[;]$ denotes row concatenation. Similarly, we write measurements as $\vy_i=[\vy_{1i};\vy_{2i};\vy_{3i};\vy_{4i}]; \vy_i\in\C^{n^2}$, where $\vy_{1i},\vy_{1i},\vy_{1i},\vy_{1i} \in\C^{n^2}$ are vectorization notation for measurements $i_{1i}(\vrr),i_{2i}(\vrr),i_{3i}(\vrr),i_{4i}(\vrr)$, and $\vp_1,\vp_2,\vp_3,\vp_4\in\C^{4n^2}$ are vectors representing pupils $P_{xx},P_{xy},P_{yx},P_{yy}$. $m^2$ and $n^2$ are pixel size of the object and camera. We can then rewrite the forward model as 

\begin{equation}
\vy_i = \abs{\mH\mP\mD_i\hat{\vx}},
\end{equation}
where $\hat{\vx}$ is the ground truth of $\vx$, $\mD_i$ is composed of four truncating matrix $\mQ_i\in \B^{n^2\times m^2}, \B=\{0,1\},$ that crops and keeps the low-passed frequency region for the $i^{th}$ illumination, and $\mH$ contains four 2D inverse Fourier transform matrix $\mF\in\C^{n^2\times n^2}$,
\newcommand\bigzero{\makebox(0,0){\text{\huge0}}}
\newcommand{\Dmatrix}{
  \begin{bmatrix}
   &\mQ_i &\mathbf{0} &\mathbf{0} &\mathbf{0} \\
   &\mathbf{0}  &\mQ_i &\mathbf{0} &\mathbf{0} \\
   &\mathbf{0} &\mathbf{0} &\mQ_i &\mathbf{0} \\
   &\mathbf{0} &\mathbf{0} &\mathbf{0} &\mQ_i
   \end{bmatrix}}
   
\newcommand{\Hmatrix}{
  \begin{bmatrix}
   &\mF &\mathbf{0} &\mathbf{0} &\mathbf{0} \\
   &\mathbf{0}  &\mF &\mathbf{0} &\mathbf{0} \\
   &\mathbf{0} &\mathbf{0} &\mF &\mathbf{0} \\
   &\mathbf{0} &\mathbf{0} &\mathbf{0} &\mF
   \end{bmatrix}}
  
\begin{equation}
   \mD_i = \Dmatrix, \mH = \Hmatrix;
\end{equation}
$\mathbf{0}$ is a all-zero matrix. $\mP$ is the pupil matrix
\providecommand{\diag}{\text{Diag}}
\newcommand{\PPmatrix}{
  \begin{bmatrix}
   &\diag(\vp_1) &\mathbf{0} &\diag(\vp_2) &\mathbf{0} \\
   &\mathbf{0} &\diag(\vp_3) &\mathbf{0} &\diag(\vp_4) \\
   &\diag(\vp_1) &\diag(\vp_1) &\diag(\vp_2) &\diag(\vp_2) \\
   &\diag(\vp_3) &\diag(\vp_3) &\diag(\vp_4) &\diag(\vp_4) 
   \end{bmatrix}}

\begin{equation}
    \mP=\PPmatrix,
\end{equation}
where $\diag(\mathbf{a})$ represents a diagonal matrix with the vector $\mathbf{a}$ on its diagonal. We aim to recover $\hat{\vx}$ from a sequence of $L$ measurements $y_i$ from $i^{th}$ LED, by minimizing the Euclidean distance between predicted and measured amplitude, as an amplitude-based loss gives empirically improved results comparing to intensity-based loss\cite{yeh2015experimental}, 
\begin{equation}
    f(\vx) = \sum_{i=1}^{L}\norm{\abs{\vzz_i}-\vy_i}_2^2,
    \label{eq::loss_fnc}
\end{equation}
where 
\begin{equation}
    \vzz_i\coloneqq\mH\mP\mD_i\vx
\end{equation}
is the measurement with phase, and $L$ is the total LED numbers. We attempt to minimize this quadratic objective function using a sequential Gauss-Newton-based method, which minimize a part of the loss function $f_i(\vx) = \frac{1}{2}\norm{\abs{\vzz_i}-\vy_i}_2^2$ in sequence. Using the $\C\R$ - calculus\cite{kreutz2009complex} that treats $\vx$ and its complex conjugate $\bar{\vx}$ as independent variables, we can write out the Jacobian as \footnote{note that we drop the $\frac{1}{2}$ in eq. \ref{eq::loss_fnc} that is commonly used to cancel the "$\norm{\cdot}^2$" introduced by the quadratic term, thus, the Jacobian in eq.\ref{eq::jacobian} does not have to have the $\frac{1}{2}$ due to the derivative of "$\sqrt{\cdot}$"}
\begin{equation}
    \mJ_i=\frac{\partial \abs{\vzz_i}}{\partial \vx}
    = \diag(\overline{\vzz_i}/\abs{\vzz_i})\mH\mP\mD_i,.
    \label{eq::jacobian}
\end{equation}
'$/$' is the element-wise division, and the Hessian can be approximated with
\begin{equation}
    \mJ_i^\text{H}\mJ_i=\mD_i^{\text{H}}\mP^{\text{H}}\mP\mD_i.
\end{equation}
$(\cdot)^\text{H}$ is the Hermitian adjoint. Starting with $\vx^0$, the iteration proceeds as 
\begin{equation}
    \vx^{t,i+1} = \vx^{t,i} - \gamma\big(\mJ_i^\text{H}\mJ_i\big)^{-1}\mJ_i^\text{H}(\abs{\vzz_i(\vx^{t,i})}-\vy_i).
    \label{eq::gn_update}
\end{equation}
Moreover,
\newcommand{\PmatrixAdj}{
  \begin{bmatrix}
   &\diag(\vp_4/\vq) &\diag(\vp_2/\vq) &\mathbf{0} &\diag(\vp_2/\vq) \\
   &\diag(-\vp_4/\vq) &\diag(\vp_3/\vq) &\diag(\vp_4/\vq) &\mathbf{0} \\
   &\diag(\vp_1/\vq) &\diag(\vp_1/\vq) &\diag(\vp_2/\vq) &\diag(\vp_2/\vq) \\
   &\diag(\vp_3/\vq) &\diag(\vp_3/\vq) &\diag(\vp_4/\vq) &\diag(\vp_4/\vq) 
   \end{bmatrix}}
   
\begin{subequations}
\begin{align} 
\big(\mJ_i^\text{H}\mJ_i\big)^{-1}\mJ_i^\text{H}=&\mD_i^{\text{H}}\big(\mP^{\text{H}}\mP\big)^{-1}\mD_i^{\text{H}}\mD_i\mP^{\text{H}}\mH^{\text{H}}\diag(\vzz_i/\abs{\vzz_i})\\
=&\mD_i^{\text{H}}\big(\mP^{\text{H}}\mP\big)^{-1}\mP^{\text{H}}\mH^{\text{H}}\diag(\vzz_i/\abs{\vzz_i})\\
=&\mD_i^{\text{H}}\mP^{\dagger}\mH^{\text{H}}\diag(\vzz_i/\abs{\vzz_i}),
\end{align}
\end{subequations}
where
\begin{align}
    \mP^{\dagger}&\coloneqq \big(\mP^{\text{H}}\mP\big)^{-1}\mP\\
    &=\PmatrixAdj,
\end{align}
 $\vq=\vp_1\cdot\vp_4-\vp_2\cdot\vp_3$, and again, '$(/)$' and '$(\cdot)$' are the element-wise division and multiplication, respectively. It's worth mentioning that $\mP^{\dagger}$ is the left inverse of the $\mP$ for the benefit of readers who are new to the field; hence, historically it is also widely referred as an alternating projection method(see discussions in \cite{qian2014efficient}). Moreover, the adjoint of the cropping matrix, $\mD_i^{\text{H}}$, is a zero-padding matrix. Thus, the update in \ref{eq::gn_update} can be explicitly implemented as

\begin{subequations}
    \begin{empheq}[left={\empheqlbrace\,}]{align}
    O_{xx}(\vu-\vu_i)&=O_{xx}(\vu-\vu_i)-Q^*(\vu)\Psi_{1i}(\vu)/\abs{Q(\vu)}^2\\
    O_{xy}(\vu-\vu_i)&=O_{xy}(\vu-\vu_i)-Q^*(\vu)\Psi_{2i}(\vu)/\abs{Q(\vu)}^2\\
    O_{yx}(\vu-\vu_i)&=O_{yx}(\vu-\vu_i)-Q^*(\vu)\Psi_{3i}(\vu)/\abs{Q(\vu)}^2\\
    O_{yy}(\vu-\vu_i)&=O_{yy}(\vu-\vu_i)-Q^*(\vu)\Psi_{4i}(\vu)/\abs{Q(\vu)}^2,
    \end{empheq}
    \label{eq::gn_implement}
\end{subequations}
where
\begin{equation}
    Q(\vu) = P_{xx}(\vu)P_{yy}(\vu)-P_{xy}(\vu)P_{yx}(\vu),
\end{equation}

\begin{subequations}
    \begin{empheq}[left={\empheqlbrace\,}]{align}
    \Psi_1(\vu) &= P_{yy}(\vu)\Phi_{1i}(\vu)+P_{xy}(\vu)\Phi_{2i}(\vu)+P_{xy}(\vu)\Phi_{4i}(\vu)\\
    \Psi_2(\vu) &=-P_{yy}(\vu)\Phi_{1i}(\vu)+P_{yx}(\vu)\Phi_{2i}(\vu)+P_{yy}(\vu)\Phi_{3i}(\vu)\\
    \Psi_3(\vu) &=P_{xx}(\vu)\Phi_{1i}(\vu)+P_{xx}(\vu)\Phi_{2i}(\vu)+P_{xy}(\vu)\Phi_{3i}(\vu)+P_{xy}(\vu)\Phi_{4i}(\vu)\\
    \Psi_4(\vu) &=P_{yx}(\vu)\Phi_{1i}(\vu)+P_{yx}(\vu)\Phi_{2i}(\vu)+P_{yy}(\vu)\Phi_{3i}(\vu)+P_{yy}(\vu)\Phi_{4i}(\vu),
    \end{empheq}
\end{subequations}

and
\begin{equation}
    \Phi_{mi}(\vu) =  \mathcal{F}\Big\{\Big(\vzz_{mi}(\vrr)-\frac{\vzz_{mi}(\vrr)}{\abs{\vzz_{mi}(\vrr)}}\sqrt{i_{mi}(\vrr)}\Big)\Big\}, m = 1,2,3,4
\end{equation}
respectively. In addition, we introduce a $\abs{Q}_{\text{max}}^2$ regularization to the division, which was shown to be robust and stable\cite{rodenburg2004phase,maiden2017further}, and the final proposed update becomes
\begin{subequations}
    \begin{empheq}[left={\empheqlbrace\,}]{align}
    O_{xx}(\vu-\vu_i)&=O_{xx}(\vu-\vu_i)-\frac{Q^*(\vu)\Psi_{1i}(\vu)}{(1-\alpha)\abs{Q(\vu)}^2+\alpha\abs{Q(\vu)}_{\text{max}}^2}\\
    O_{xy}(\vu-\vu_i)&=O_{xy}(\vu-\vu_i)-\frac{Q^*(\vu)\Psi_{2i}(\vu)}{(1-\alpha)\abs{Q(\vu)}^2+\alpha\abs{Q(\vu)}_{\text{max}}^2}\\
    O_{yx}(\vu-\vu_i)&=O_{yx}(\vu-\vu_i)-\frac{Q^*(\vu)\Psi_{3i}(\vu)}{(1-\alpha)\abs{Q(\vu)}^2+\alpha\abs{Q(\vu)}_{\text{max}}^2}\\
    O_{yy}(\vu-\vu_i)&=O_{yy}(\vu-\vu_i)-\frac{Q^*(\vu)\Psi_{4i}(\vu)}{(1-\alpha)\abs{Q(\vu)}^2+\alpha\abs{Q(\vu)}_{\text{max}}^2}.
    \end{empheq}
\end{subequations}
Similarly, we can derive the update for the pupils
\begin{subequations}
    \begin{empheq}[left={\empheqlbrace\,}]{align}
    P_{xx}(\vu)&=P_{xx}(\vu)-\frac{T_i^*(\vu)\Lambda_{1i}(\vu)}{(1-\alpha)\abs{T_i(\vu)}^2+\alpha\abs{T_i(\vu)}_{\text{max}}^2}\\
    P_{xy}(\vu)&=P_{xy}(\vu)-\frac{T_i^*(\vu)\Lambda_{2i}(\vu)}{(1-\alpha)\abs{T_i(\vu)}^2+\alpha\abs{T_i(\vu)}_{\text{max}}^2}\\
    P_{yx}(\vu)&=P_{yx}(\vu)-\frac{T_i^*(\vu)\Lambda_{3i}(\vu)}{(1-\alpha)\abs{T_i(\vu)}^2+\alpha\abs{T_i(\vu)}_{\text{max}}^2}\\
    P_{yy}(\vu)&=P_{yy}(\vu)-\frac{T_i^*(\vu)\Lambda_{4i}(\vu)}{(1-\alpha)\abs{T_i(\vu)}^2+\alpha\abs{T_i(\vu)}_{\text{max}}^2},
    \end{empheq}
\end{subequations}
where
\begin{equation}
    T_i = O_{xx}(\vu-\vu_i)O_{yy}(\vu-\vu_i)-O_{xy}(\vu-\vu_i)O_{yx}(\vu-\vu_i),
\end{equation}
and
\begin{subequations}
    \begin{empheq}[left={\empheqlbrace\,}]{align}
    \Lambda_{1i}(\vu)&=O_{yx}(\vu-\vu_i)\Phi_{2i}(\vu)+O_{yy}(\vu-\vu_i)\Phi_{2i}(\vu)-O_{yx}(\vu-\vu_i)\Phi_{3i}(\vu)\\
    \Lambda_{2i}(\vu)&=-O_{xx}(\vu-\vu_i)\Phi_{2i}(\vu)-O_{xy}(\vu-\vu_i)\Phi_{2i}(\vu)+O_{xx}(\vu-\vu_i)\Phi_{3i}(\vu)\\
    \Lambda_{3i}(\vu)&=-O_{yx}(\vu-\vu_i)\Phi_{1i}(\vu)-O_{yy}(\vu-\vu_i)\Phi_{1i}(\vu)+O_{yy}(\vu-\vu_i)\Phi_{4i}(\vu)\\
    \Lambda_{4i}(\vu)&=O_{xx}(\vu-\vu_i)\Phi_{1i}(\vu)+O_{xy}(\vu-\vu_i)\Phi_{1i}(\vu)-O_{xy}(\vu-\vu_i)\Phi_{4i}(\vu)
    \end{empheq}
\end{subequations}

\section{Analyzing polarization state with a linear polarizer}
\begin{figure}
\begin{center}
  \includegraphics[width=13 cm]{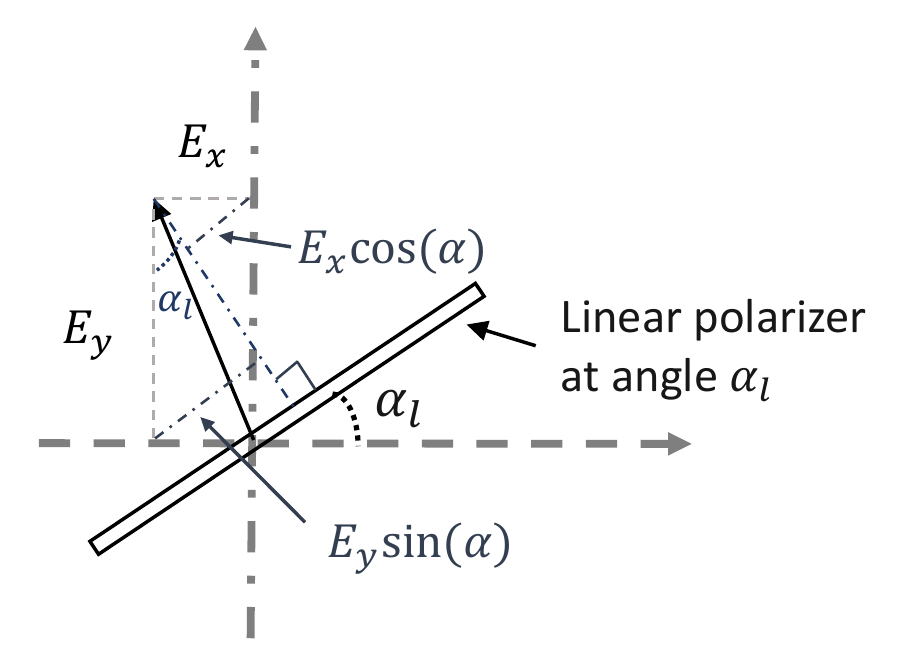} 
   \caption{Graphic explaination of eq.\ref{eq::analyz}. The slit is the linear polarizer. $\alpha$ is the angle between the linear polarizer and the x-polarization. Projecting the $E_x$ and $E_y$ components of the incoming electric field to the polarizer direction yields a passing electric field $E_x\text{cos}(\alpha_l)+E_y\text{sin}(\alpha_l)$ }
\label{analyzer}
\label{fig::analyzer}
\end{center}
\end{figure}

Here we derive a simple way to compute the detected intensity value after the electric field $\mathbf{e}=[E_x,E_y]^T$ passing through the $l^{th}$ analyzer, made with a linear polarizer placed at angle $\alpha_l$ with respect to the x-polarization. This handy equation is used in the eq(3) in the main text. The Jones matrix for the analyzer is \cite{chipman2018polarized}
\begin{equation}
    \mathbf{M}_{\alpha_l}=\begin{bmatrix}
\text{cos}^2(\alpha_l) & \text{cos}(\alpha_l)\text{sin}(\alpha_l)\\
\text{cos}(\alpha_l)\text{sin}(\alpha_l) & \text{sin}^2(\alpha_l)
\end{bmatrix}.
\end{equation}
Hence,  the electric field exiting the polarizer is 
\begin{align}
    &\mathbf{e}'= \mathbf{M}_{\alpha_l}\mathbf{e}\\
    &\mkern-36mu=\begin{bmatrix}
     \text{cos}^2(\alpha_l)E_x+\text{cos}(\alpha_l)\text{sin}(\alpha_l)E_y
     \\ \text{cos}(\alpha_l)\text{sin}(\alpha_l)E_x+\text{sin}^2(\alpha_l)E_y
    \end{bmatrix}.
\end{align}
Similarly, we denote the x-y polarization components of the existing electric field $\mathbf{e}'=[E_x',E_y']^T$.  After a bit tedious math, the detected intensity can be calculated as 
\begin{align}
    i_{\alpha_l}=&\abs{\mathbf{e}'}^2\\
    =&\abs{E_x'}^2+\abs{E_y'}^2\\
    =&\abs{\text{cos}\alpha_l E_x+\text{sin}\alpha_l E_y}^2\\
    =&\abs{\textbf{a}_l^T\textbf{e}}^2,
\label{eq::analyz}
\end{align}
where $\mathbf{a}_l=[\text{cos}(\alpha_l),\text{sin}(\alpha_l)]^T$. In addition, an alternative intuitive graphic explanation is also provided as fig. \ref{fig::analyzer}

\end{document}